\documentclass[aps,pra,reprint,groupedaddress,superscriptaddress,floatfix]{revtex4-2}

\usepackage[T1]{fontenc}
\usepackage{newtxtext}
\usepackage{newtxmath}

\usepackage{enumitem}

\usepackage{amsmath}
\usepackage{mathtools}
\usepackage{amsfonts}

\usepackage{color}
\usepackage{xcolor}
\usepackage[colorlinks, citecolor=blue, linkcolor=blue, urlcolor=blue]{hyperref}

\usepackage{graphicx}
\usepackage[all]{hypcap}

\begin{document}

\title{An entanglement perspective on the quantum approximate optimization algorithm}

\author{Maxime Dupont}
\email[Corresponding author: ]{mdupont@rigetti.com}
\affiliation{Department of Physics, University of California, Berkeley, California 94720, USA}
\affiliation{Materials Sciences Division, Lawrence Berkeley National Laboratory, Berkeley, California 94720, USA}
\affiliation{Rigetti Computing, 775 Heinz Avenue, Berkeley, California 94710, USA}

\author{Nicolas Didier}
\affiliation{Rigetti Computing, 775 Heinz Avenue, Berkeley, California 94710, USA}

\author{Mark J. Hodson}
\affiliation{Rigetti Computing, 775 Heinz Avenue, Berkeley, California 94710, USA}

\author{Joel E. Moore}
\affiliation{Department of Physics, University of California, Berkeley, California 94720, USA}
\affiliation{Materials Sciences Division, Lawrence Berkeley National Laboratory, Berkeley, California 94720, USA}

\author{Matthew J. Reagor}
\affiliation{Rigetti Computing, 775 Heinz Avenue, Berkeley, California 94710, USA}

\begin{abstract}
    Many quantum algorithms seek to output a specific bitstring solving the problem of interest---or a few if the solution is degenerate. It is the case for the quantum approximate optimization algorithm (QAOA) in the limit of large circuit depth, which aims to solve quadratic unconstrained binary optimization problems. Hence, the expected final state for these algorithms is either a product state or a low-entangled superposition involving a few bitstrings. What happens in between the initial $N$-qubit product state $\vert 0\rangle^{\otimes N}$ and the final one regarding entanglement? Here, we consider the QAOA algorithm for solving the paradigmatic MaxCut problem on different types of graphs. We study the entanglement growth and spread resulting from randomized and optimized QAOA circuits and find that there is a volume-law entanglement barrier between the initial and final states. We also investigate the entanglement spectrum in connection with random matrix theory. In addition, we compare the entanglement production with a quantum annealing protocol aiming to solve the same MaxCut problems. Finally, we discuss the implications of our results for the simulation of QAOA circuits with tensor network-based methods relying on low-entanglement for efficiency, such as matrix product states.
\end{abstract}

\maketitle

\section{Introduction}

Entanglement is an essential component of quantum mechanics that makes quantum computers fundamentally different from their classical counterparts. Quantum algorithms leverage the ability to generate entangled quantum superpositions of states to bring quantum speedup. Two celebrated examples are Shor's algorithm for factoring integers~\cite{Shor1994} and Grover's for unstructured search~\cite{Grover1996}. Therefore, investigating quantum algorithms through the prism of entanglement may provide valuable information.

Indeed, entanglement is routinely used to characterize properties of quantum many-body states, see, e.g., Ref.~\cite{LAFLORENCIE20161} for a review. By tracing out a subset of the degrees of freedom, one obtains the reduced density matrix describing the remaining subsystem. Its eigenvalue spectrum and corresponding R\'enyi entropies contain key information on the system. They can help diagnose symmetry breaking, chaos, topological, and localization features, to cite but a few. Looking at entanglement growth and spread following a quantum quench on a product state can help understand how quantum information settles in the system, in connection, e.g, with thermalization and relaxation to equilibrium. This is precisely this last point that is of interest in the context of quantum computing: How does entanglement spread following consecutive layers of unitary gates applied on the initial $N$-qubit state $\vert 0\rangle^{\otimes N}$? In addition and in connection with nonequilibrium quantum many-body physics, entanglement growth in random circuits is subject to extensive research~\cite{PhysRevX.7.031016,PhysRevX.8.031058,Znidaric2020}; see also Ref.~\cite{Fisher2022} for a recent review article.

Moreover, entanglement plays a critical role in the success of classical approximate quantum simulators based on tensor networks such as matrix product states (MPSs)~\cite{PhysRevLett.93.040502,Schollwock2011,Orus2014}, projected entangled-pair states~\cite{Verstraete2004,PhysRevLett.96.220601}, tree tensor networks~\cite{PhysRevA.74.022320}, the multi-scale entanglement renormalization ansatz~\cite{PhysRevLett.101.110501}, and two-dimensional isometric tensor networks~\cite{PhysRevLett.124.037201}. They all have a control parameter $\chi$, the so-called bond dimension, which governs the allocated classical computational power where time complexity and memory usage scale polynomially with $\chi$. Besides managing resources, $\chi$ has a physical interpretation: It dictates the amount of entanglement that can be encoded in the tensor network, i.e., $S\sim\ln\chi$ for MPSs~\cite{PhysRevLett.93.040502,Schollwock2011,Orus2014}. While approximate simulators can accommodate exact calculations if the bond dimension scales exponentially with the number of qubits, i.e., $\chi_\mathrm{exact}\equiv\chi=2^{N/2}$ for MPSs, there are situations in which the parameter needs to be effectively smaller than that. For instance, many-body states in one dimension fulfilling the area law contain finite entanglement $S\sim\mathrm{const}$~\cite{Hastings2007,Eisert2010}, which means they can be represented accurately as MPSs with finite $\chi$, independently of $N$. Celebrated examples include the ground state of gapped Hamiltonians and (many-body) localized eigenstates at arbitrary energy. Studying the entanglement generated by quantum circuits can help us understand the limitations of approximate simulators relying on low-entangled quantum states: How does the entanglement scale versus the number of qubits $N$? In turn, it sets the scaling of the bond dimension $\chi$ with $N$. In regard to quantum computing, MPSs have been used to simulate Shor's algorithm~\cite{Shor1994,Wang2017,Dang2019}, Google's Sycamore circuits~\cite{Oh2021}, boson sampling~\cite{Zhou2020}, and the quantum approximate optimization algorithm (QAOA)~\cite{Farhi2014,Farhi2014b,Farhi2016,Patti2021,Dupont2022}.

QAOA belongs to the more general class of variational quantum algorithms where one optimizes a parametric quantum circuit such that its output minimizes a given cost function~\cite{Cerezo2021}. It is designed to solve quadratic unconstrained binary optimization problems~\cite{Kochenberger2014}, with applications in logistics, scheduling, finance, traffic congestion, machine learning, and basic science among others. It is largely regarded by the community as a promising candidate for delivering a quantum advantage to real-world problems in the near term and thus subject to extensive research~\cite{Farhi2014,Farhi2014b,Farhi2016,Wecker2016,Guerreschi2017,Otterbach2017,Jiang2017,Verdon2017,Lloyd2018,Crooks2018,Qiang2018,PhysRevA.97.022304,Lechner2018,Anschuetz2018,Sundar2019,Hastings2019,Hadfield2019,Guerreschi2019,Lykov2020,Dalzell2020,Bravyi2020,ZhouLeo2020,Wiersema2020,Wierichs2020,Pagano2020,Bengtsson2020,Wang2020,Willsch2020,Abrams2020,Patti2021,Medvidovic2021,Kremenetski2021,Barak2021,Fitzek2021,Juneseo2021,Dumitrescu2018,Herrman2021,Akshay2021,Harrigan2021,DiezValle2022,Zhu2022,Sohaib2022,Ebadi2022,Stollenwerk2022,Amaro2022,Lotshaw2022,GonzalezGarcia2022,Weidenfeller2022,Santra2022,Chen2022}.

Reference~\cite{Dupont2022} recently considered MPS simulations of the QAOA algorithm. The authors explored how working with a value of $\chi$ smaller than the one required for exact simulations modified the overall quality of the circuit execution. They assessed the output fidelity through the expectation value of the cost function QAOA seeks to minimize and found that a small bond dimension makes the cost larger than in the ideal case. Because $\ln\chi$ amounts for the maximum entanglement that can be encoded by an MPS, the results suggest that QAOA circuits generate a lot of entanglement that was being cut off by restricting $\chi$.

\begin{figure}[t]
    \includegraphics[width=1\columnwidth]{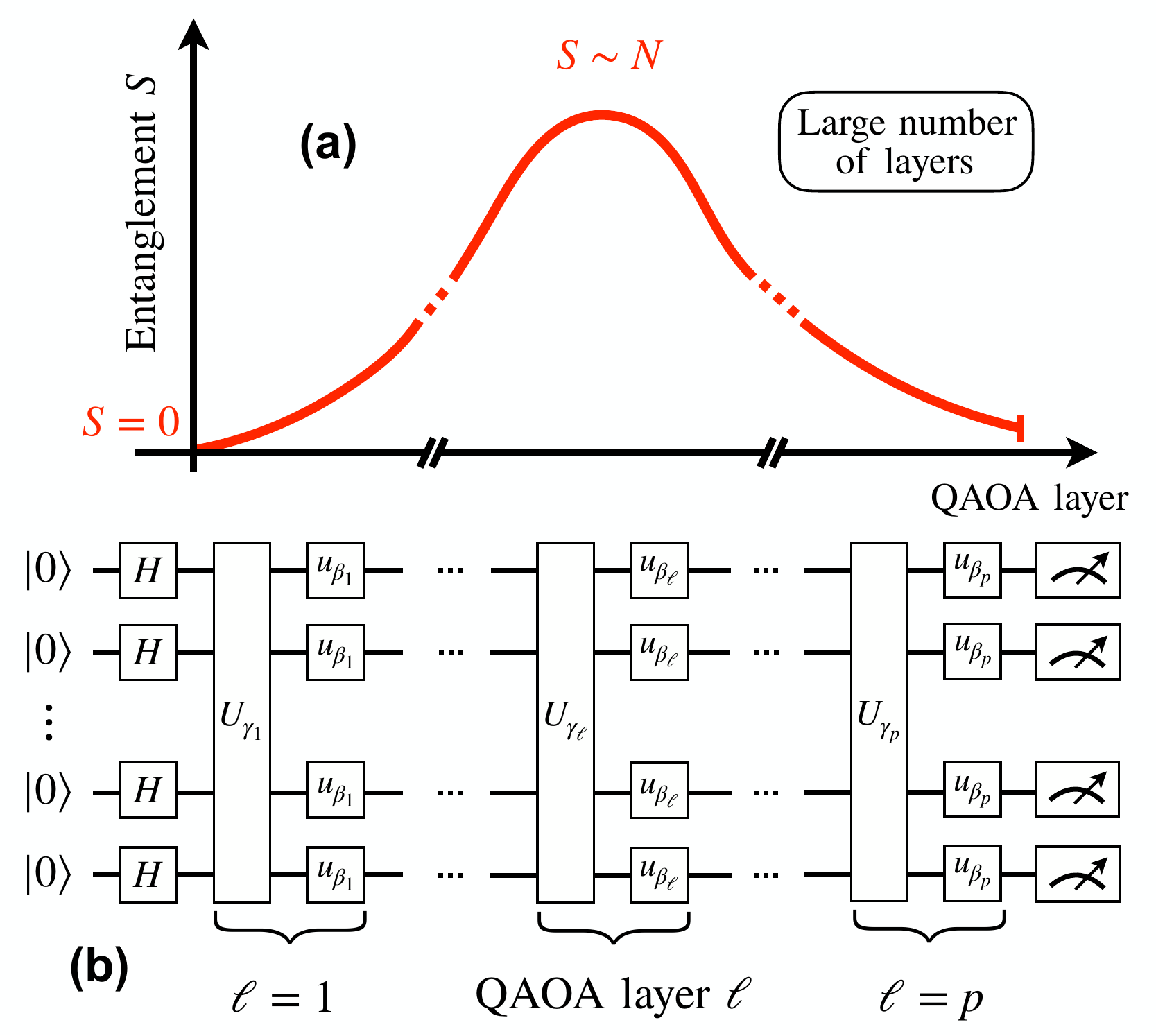} 
    \caption{(a) Sketch of the bipartite entanglement entropy $S$ as a function of the QAOA layer for solving a generic MaxCut problem. The initial state is a product state with $S=0$. In the limit of a large number of layers, QAOA solves asymptotically the combinatorial problem of interest. Thus one expects the final state to be low-entangled. In intermediate steps, optimized QAOA circuits generate volume-law entanglement $S\sim N$ with $N$ the number of qubits. (b) A QAOA circuit according to Eq.~\eqref{eq:qaoa_circuit} with layers $\ell\leq p$. We used $U_{\beta_\ell}=\prod\nolimits_ju_{\beta_\ell}$ with $u_{\beta_\ell}$ acting on individual qubits according to Eq.~\eqref{eq:qaoa_unitaries}.}
    \label{fig:introduction}
\end{figure}

Here, we get an entanglement perspective on QAOA circuits. In the limit of large circuit depth, QAOA converges asymptotically to the few bitstrings solving the optimization problem of interest. Hence, the final quantum state is a low-entangled superposition, which can be represented accurately as an MPS with finite $\chi$. What happens in between the initial and final low-entangled states? We find that there is a volume-law entanglement barrier which requires $\chi\sim\exp(N)$, see Fig.~\ref{fig:introduction}(a). Precisely, we study the bipartite von Neumann entanglement entropy and the entanglement spectrum resulting from QAOA circuits for the paradigmatic MaxCut problem. We consider randomized QAOA circuits as well as optimized QAOA circuits.

The rest of the paper is organized as follows. We present definitions and methods in Sec.~\ref{sec:def_methods}. In Sec.~\ref{sec:ent_random}, we investigate entanglement in randomized QAOA circuits, which is a typical starting point for variational quantum algorithms. We repeat the analysis in Sec.~\ref{sec:ent_opt} for optimized QAOA circuits. In Sec.~\ref{sec:ent_annealing}, we compare the entanglement in QAOA circuits with a quantum annealing protocol where the circuit parameters are fixed by the interpolation towards the cost function that one seeks to minimize. Finally, we summarize our results and their implications in Sec.~\ref{sec:conclusion}.

\section{Definitions and Methods}
\label{sec:def_methods}

\subsection{The MaxCut Problem}

Given an undirected graph $G=(V, E)$ with vertex set $V$ and edge set $E$, where each edge $\{i,j\}\in E$ has a weight $w_{ij}\geq 0$, the MaxCut problem seeks to find a cut maximizing the total weight of cut edges. The problem is NP-hard, with no known polynomial-time algorithms that will return the maximum cut for general graphs.

Assigning a variable $s_i=\pm 1$ to each vertex $i\in V$ of the graph, the maximum cut corresponds to the configuration $\boldsymbol{s}=(s_1,s_2\ldots s_N)$ minimizing~\cite{Lucas2014},
\begin{equation}
    C(\boldsymbol{s})=\sum_{\{i,j\}\in E}w_{ij}s_is_j,
    \label{eq:MaxCut_cost}
\end{equation}
with the edge $\{i,j\}\in E$ being cut if $s_i\neq s_j$. In other words, minimizing Eq.~\eqref{eq:MaxCut_cost} is analogous to finding the two complementary sets of vertices $\{-1\}$ and $\{+1\}$ such that the total weight of the edges between them is maximized. A configuration with all the signs flipped, i.e., $\boldsymbol{s}\to-\boldsymbol{s}$, corresponds to the same cut.

Although solving the MaxCut problem is NP-hard, there exists efficient polynomial-time algorithms that find an approximate solution $\boldsymbol{s}_\mathrm{approx}$ whose cost is at least as low as $rC(\boldsymbol{s}_\mathrm{exact})$ with high probability, where $\boldsymbol{s}_\mathrm{exact}$ is the exact solution and $r\leq 1$ the approximation ratio. While it remains NP-hard to find a solution with $r\geq 16/17\simeq 0.941$~\cite{Hastad2001}, the Goemans-Williamson algorithm~\cite{Goemans1995} guarantees, for instance, an approximation ratio of $r=0.878$ for general graphs, and which can be improved in certain cases~\cite{Halperin2004}.

\subsection{The Quantum Approximate Optimization Algorithm}

The QAOA circuit on $N$ qubits with $p$ layers and $2p$ parameters is as follows~\cite{Farhi2014,Farhi2014b,Farhi2016},
\begin{equation}
    \bigl\vert\boldsymbol{\theta}\equiv\bigl\{\boldsymbol{\beta},\boldsymbol{\gamma}\bigr\}\bigr\rangle=\Biggr(\prod_{\ell=1}^pU_{\beta_\ell}U_{\gamma_\ell}\Biggl)H^{\otimes N}\vert 0\rangle^{\otimes N},
    \label{eq:qaoa_circuit}
\end{equation}
with $H$ the Hadamard gate applied on the individual qubits, where each of them corresponds to a vertex of the graph. The parametrized unitaries are,
\begin{equation}
    U_{\beta_\ell}=\prod\nolimits_j\exp\Biggl(-i\frac{\beta_\ell}{2}X_j\Biggr)~\mathrm{and}\quad U_{\gamma_\ell}=\exp\Biggl(-i\frac{\gamma_\ell}{2}C\Biggr),
    \label{eq:qaoa_unitaries}
\end{equation}
with $X_j$ the Pauli operator on qubit $j$ and $C$ the cost function operator related to the MaxCut problem on a graph $G=(V,E)$. It is diagonal in the computational basis,
\begin{equation}
    C=\sum_{\{i,j\}\in E}w_{ij}Z_iZ_j,
    \label{eq:qaoa_cost}
\end{equation}
with the sum running over the edges $\{i,j\}$ according to Eq.~\eqref{eq:MaxCut_cost}. $Z_j$ is the Pauli operator on qubit $j$. The unitary $U_{\beta}$ is a collection of single-qubit rotations about the $x$ axis of the Bloch sphere. On a gate-based quantum computer, the global unitary $U_{\gamma}$ can be broken down into two-qubit gates acting on nearest-neighbor qubits as per the underlying graph topology. A circuit with layers $\ell\leq p$ is displayed in Fig.~\ref{fig:introduction}(b).

Given the parametric circuit of Eq.~\eqref{eq:qaoa_circuit}, the goal is then to classically optimize its parameters for the circuit output to minimize the cost function of the MaxCut problem,
\begin{equation}
    C_\mathrm{min}=\min_{\boldsymbol{\theta}}~\bigl\langle\boldsymbol{\theta}\bigr\vert C\bigl\vert\boldsymbol{\theta}\bigr\rangle,
    \label{eq:qaoa_cost_expectation}
\end{equation}
which can be achieved by traditional optimization algorithms.

\subsection{Graphs}

\begin{figure}[t]
    \includegraphics[width=0.65\columnwidth]{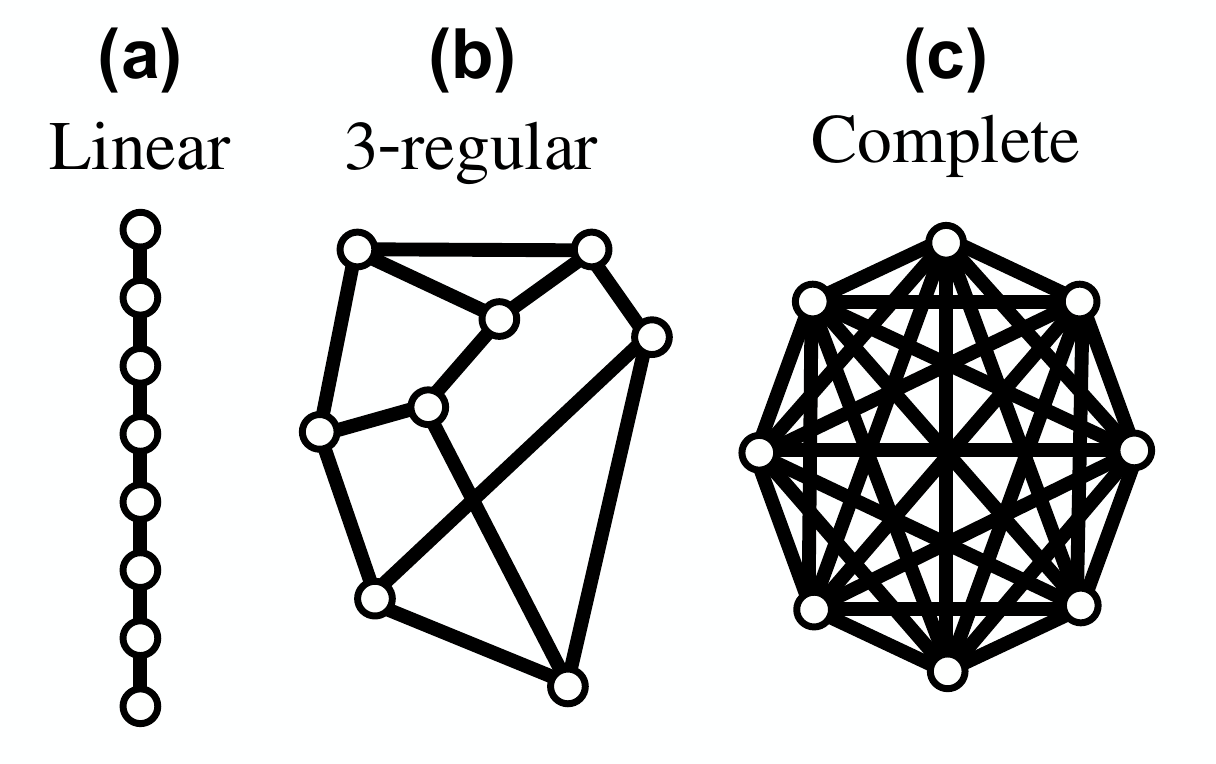} 
    \caption{Sketch of the graphs considered in this paper with $N=8$ vertices. (a) Linear graph. (b) A $3$-regular graph. (c) Complete graph.}
    \label{fig:graphs}
\end{figure}

We investigate QAOA for the MaxCut problem on three types of graphs:
\begin{enumerate}[label=(\arabic*),leftmargin=15pt,itemsep=-3pt]
    \item Linear graphs: First, we consider linear graphs with $N$ vertices where the first and last vertices are not connected. Each of the $N-1$ edges carry a random weight $w_{ij}\in[0,1]$. See Fig.~\ref{fig:graphs}(a).
    \item 3-regular graphs: Then, we consider $3$-regular graphs with $N$ vertices where each vertex is connected to three other and distinct vertices at random (there are no self-loops or parallel edges). The graphs are uniform with unit weight $w_{ij}=1$. See Fig.~\ref{fig:graphs}(b).
    \item Complete graphs: Finally, we consider complete graphs (all-to-all connected) with $N$ vertices and assign a random weight $w_{ij}\in[0,1]$ to the edges. See Fig.~\ref{fig:graphs}(c).
\end{enumerate}
This choice is guided by the increased connectivity of the graphs, which should favor entanglement spreading within the system with each QAOA layer, which we seek to characterize.

\subsection{Entanglement Spectrum and Entropy}

The entanglement entropy quantifies the degree of quantum entanglement between two subsets of qubits $A$ and $B$ of a system defined over $A\cup B$. For a pure state $\vert\boldsymbol{\theta}\rangle$, the reduced density matrix $\rho_A=\mathrm{tr}_{B}\vert\boldsymbol{\theta}\rangle\langle\boldsymbol{\theta}\vert$ of the subsystem $A$ (respectively, $B$) associated with such a bipartition can be used to compute the bipartite R\'enyi entropy of index $q$ between $A$ and $B$, $S_q=\ln(\mathrm{tr}\rho_A^q)/(1-q)$. In the limit $q\to 1$, the R\'enyi entropy approaches the bipartite von Neumann entanglement entropy,
\begin{equation}
    S = -\mathrm{tr}\left(\rho_A\ln\rho_A\right)=-\sum\nolimits_k\lambda_k^2\ln\lambda_k^2,
    \label{eq:entanglement_def}
\end{equation}
with $\{\lambda_k^2\}$ the eigenvalues of $\rho_A$ known as the entanglement spectrum, and which fulfill $\sum_k\lambda_k^2=1$. We consider the bipartite von Neumann entanglement entropy of Eq.~\eqref{eq:entanglement_def} in the following.

\subsection{Implementation}

We perform the QAOA simulations using a state vector approach, where $\vert\boldsymbol{\theta}\rangle$ from Eq.~\eqref{eq:qaoa_circuit} is evaluated exactly. We obtain results up to $N=22$, corresponding to an Hilbert space dimension $2^N$. We do not make use of the $\mathbb{Z}_2$ global qubit inversion symmetry $0\leftrightarrow 1$ of Eq.~\eqref{eq:qaoa_circuit}, which would otherwise reduce the size of the Hilbert space by a factor of $2$.

The minimization is carried out using the Broyden-Fletcher-Goldfarb-Shanno (BFGS) algorithm~\cite{BFGS1,BFGS2,BFGS3,BFGS4}. For each problem considered, we repeat the optimization procedure for $\approx 10^3$ random initialization of the parameters and only keep the best result. The initial $2p$ parameters are drawn from the uniform distributions: $\gamma\in[0,2\pi)$ and $\beta\in[0,\pi)$ but the optimization is unbounded. Two cases are considered: randomized QAOA circuits with random parameters in Sec.~\ref{sec:ent_random} and optimized QAOA circuits where parameters have been optimized to minimize the desired cost function in Sec.~\ref{sec:ent_opt}.

When looking at the entanglement properties, the bipartition is taken over two subsets of equal size $N/2$. On complete and $3$-regular graphs, the qubits entering $A$ and $B$ are randomly chosen as there is no natural bipartition. On the other hand, on linear graphs, we take $A$ and $B$ as two joint subsets corresponding to the two halves of the graph. Entanglement properties are computed from the Schmidt decomposition of the state $\vert\boldsymbol{\theta}\rangle$. Noting the basis states $\{\vert\boldsymbol{s}\rangle\equiv\vert\boldsymbol{s}_A\rangle\otimes\vert\boldsymbol{s}_B\rangle\}$ with $\vert\boldsymbol{\theta}\rangle=\sum_{\{\boldsymbol{s}\}}c_{\boldsymbol{s}}\vert\boldsymbol{s}\rangle$, we construct the matrix with entries $M_{\boldsymbol{s}_A\boldsymbol{s}_B}=c_{\boldsymbol{s}}$ and perform a singular value decomposition. The resulting singular values $\{\lambda_k\}$ are the Schmidt coefficients entering Eq.~\eqref{eq:entanglement_def}.

\section{Entanglement in Randomized QAOA Circuits}
\label{sec:ent_random}

\begin{figure}[t]
    \includegraphics[width=1\columnwidth]{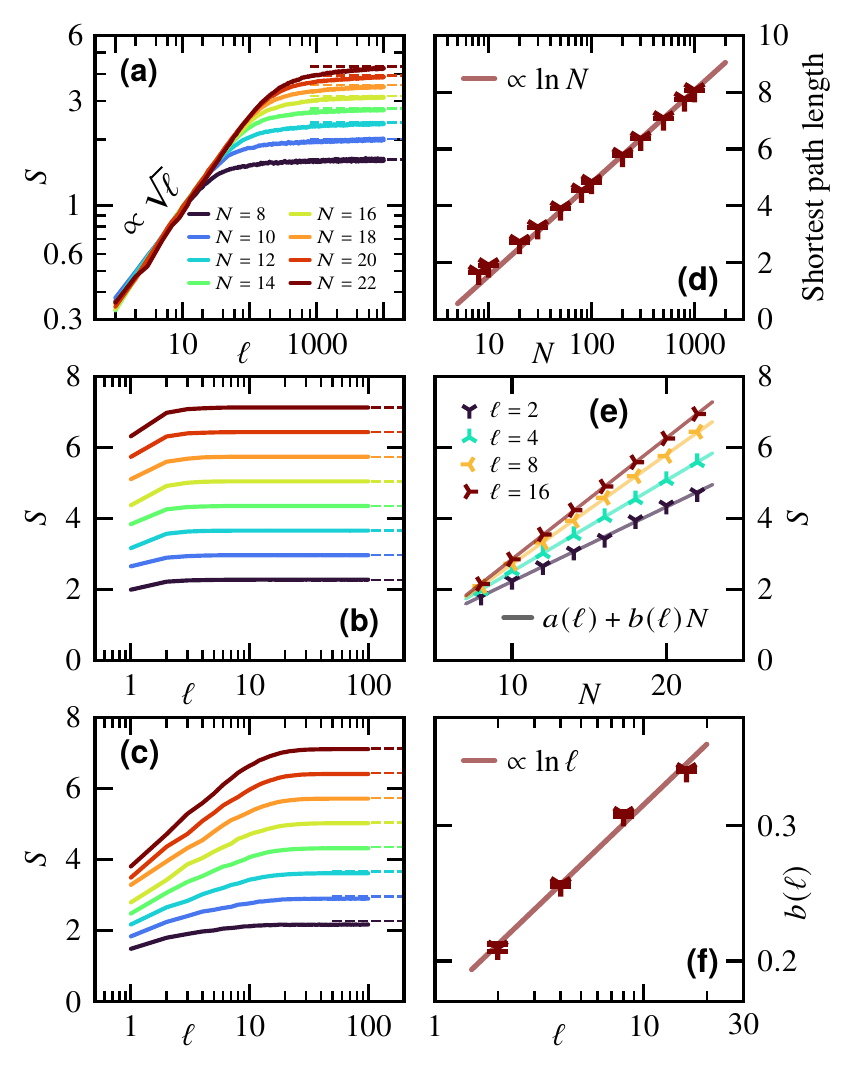} 
    \caption{Data averaged over $10^3$ problems and randomized QAOA circuits. The statistical error bars are smaller than the symbols and not displayed. (a)--(c) Growth of the bipartite von Neumann entanglement entropy $S$ defined in Eq.~\eqref{eq:entanglement_def} versus the number of randomized QAOA layers $\ell$. (a) Linear graphs. (b) Complete graphs. (c) $3$-regular graphs. The straight dashed line corresponds to the saturation value following Eq.~\eqref{eq:entanglement_saturation}. (a) For $\ell\lesssim N^2$, the entanglement in the linear graphs shows a diffusive growth $\propto\sqrt{\ell}$ with a size-independent prefactor. (b) In complete graphs, the saturation happens after a number of QAOA layers of order one. (c) Based on the properties of $3$-regular graphs, we expect the entanglement to saturate for $\ell_\mathrm{sat}\sim\ln N$, corresponding to the average shortest path between two vertices. (d) Average shortest path length in $3$-regular graphs as a function of the graph size $N$. Each data point is averaged over $10^3$ random graphs. For $N\to+\infty$, we observe that the average shortest path grows as $\sim\ln N$. For smaller graphs, the average shortest path length is larger than the genuine thermodynamic behavior. (e) For $3$-regular graphs, at fixed $\ell$, we show the entanglement growth as a function of the size $N$. Straight lines are linear fits of the form $a(\ell)+b(\ell)N$ with $a$ and $b$ $\ell$-dependent fitting parameters for $\ell\lesssim\ell_\mathrm{sat}$. (f) Fitting parameter $b(\ell)$ showing a behavior compatible with a logarithmic dependence.}
    \label{fig:random_QAOA_S}
\end{figure}

Although strategies have been proposed to find a good initial guess for the QAOA parameters $\boldsymbol{\theta}$~\cite{ZhouLeo2020}---thus accelerating the search of the global optimum---a standard starting point is randomized parameters. Therefore, studying the role of entanglement in randomized QAOA circuits is relevant in evaluating the limitations of entanglement-based classical simulation methods.

\subsection{Entanglement Entropy}

\subsubsection{Entanglement saturation}

We first consider the entanglement entropy $S$, as defined in Eq.~\eqref{eq:entanglement_def}, as a function of the number of randomized QAOA layers $\ell$ for increasing system size $N$. The data is averaged over $10^3$ problems. Results are displayed in Figs.~\ref{fig:random_QAOA_S}(a)--(c) for the three different graphs considered. In all cases, beyond a threshold value $\ell\gtrsim\ell_\mathrm{sat}$, the bipartite von Neumann entanglement entropy saturates to a volume law,
\begin{equation}
    S\bigl(N,\ell\gtrsim\ell_\mathrm{sat}\bigr) = s_0N + s_1 + O\bigl(N^{-1}\bigr),
    \label{eq:entanglement_saturation}
\end{equation}
with $s_0$ and $s_1$ universal constants. On complete and $3$-regular graphs, one recovers the Page value~\cite{PhysRevLett.71.1291} for random states, where $s_0=\ln 2/2$ and $s_1=-1/2$. For the linear graph, we find values compatible with $s_0\approx 0.193$ and $s_1\approx 0.077$, corresponding to the average entanglement entropy of a random Gaussian state~\cite{PhysRevLett.116.030401,PhysRevB.97.245126,Dias2021,PhysRevB.103.L241118}.

This difference is specific to linear graphs, where the QAOA unitaries of Eq.~\eqref{eq:qaoa_unitaries} are those of the time evolution of a one-dimensional transverse field Ising model, which maps to a free fermionic model by a Jordan-Wigner transformation. In that case, randomized QAOA circuits enter the category of random free fermionic circuits~\cite{Dias2021}. The mapping makes the classical simulation of QAOA circuits on linear graphs possible in polynomial time with the number of qubits rather than exponential~\cite{PhysRevA.65.032325,PhysRevA.97.022304,PhysRevA.104.062614}.

\subsubsection{Entanglement growth}

\paragraph{Linear graphs.} Because of the special nature of randomized QAOA circuits on linear graphs, the entanglement growth before saturation is diffusive with the number of layers~\cite{Dias2021}, i.e., $S\bigl(N,\ell\lesssim\ell_\mathrm{sat}\bigr)\sim\sqrt{\ell}$, with a system-size independent prefactor, see Fig.~\ref{fig:random_QAOA_S}(a). It is much slower than a linear growth, which is the maximum possible growth rate in a one-dimensional system with short-range gates. The square root scaling, in combination with the volume law for entanglement saturation, leads to a number of layers required for saturation scaling with the system size as $\ell_\mathrm{sat}\sim N^2$.

\paragraph{Complete graphs.} On complete graphs, see Fig.~\ref{fig:random_QAOA_S}(b), the entanglement saturates extremely quickly, after a number of QAOA layers of order one. This is understood by the graph being all-to-all connected, making all qubits aware of each others after one layer, thus favoring entanglement spreading.

\paragraph{$3$-regular graphs.} Based on the topological properties of $3$-regular graphs in the limit $N\to+\infty$, with the average shortest path between two vertices scaling with the graph size as $\sim\ln N$, see Fig.~\ref{fig:random_QAOA_S}(e), we expect the entanglement entropy to saturate for $\ell_\mathrm{sat}\sim\ln N$. Because we work with graphs $N\leq 22$ due to the exponential size of the Hilbert space with $N$, it is not possible to verify this in practice. Instead, we observe in Fig.~\ref{fig:random_QAOA_S}(c) that the entanglement saturates at what seems to be a size-independent depth $\ell_\mathrm{sat}\approx 20$. Before reaching saturation, the growth looks linear on a semi-log $x$ scale, meaning that the data suggests $S\bigl(N,\ell\lesssim\ell_\mathrm{sat}\bigr)\sim\ln\ell$. The entanglement growth before reaching saturation is analyzed further in Fig.~\ref{fig:random_QAOA_S}(e) by looking at the entanglement versus $N$ for fixed QAOA circuit depths $\ell$. Although for $\ell\lesssim\ell_\mathrm{sat}$ the entanglement entropy is not saturated, it still shows a volume law scaling, i.e., $S\bigl(N,\ell\lesssim\ell_\mathrm{sat}\bigr)=a(\ell)+b(\ell)N$ with $a$ and $b$ two constants. For the small sizes available, we find in Fig.~\ref{fig:random_QAOA_S}(f) that $b(\ell)\sim\ln\ell$, in agreement with the logarithmic growth of Fig.~\ref{fig:random_QAOA_S}(c). However, we do not believe this is the correct $N\to+\infty$ behavior, as it would mean $\ell_\mathrm{sat}\sim\mathrm{const}$, which is incompatible with the topological properties of the graph. We attribute this inconsistency to finite-size effects and expect that for larger system sizes $N$, the entanglement will grow slower. To justify this as a finite-size effect, we see in Fig.~\ref{fig:random_QAOA_S}(d) that the $\sim \ln N$ behavior for the average shortest path length in $3$-regular graphs is recovered at large $N$. For smaller sizes ($N\lesssim 50$), the average shortest path is larger than the expected genuine behavior $\sim \ln N$, potentially explaining why entanglement spreads faster than expected in the small $3$-regular graphs accessible.

\subsection{Entanglement Spectrum}

\begin{figure}[t]
    \includegraphics[width=1\columnwidth]{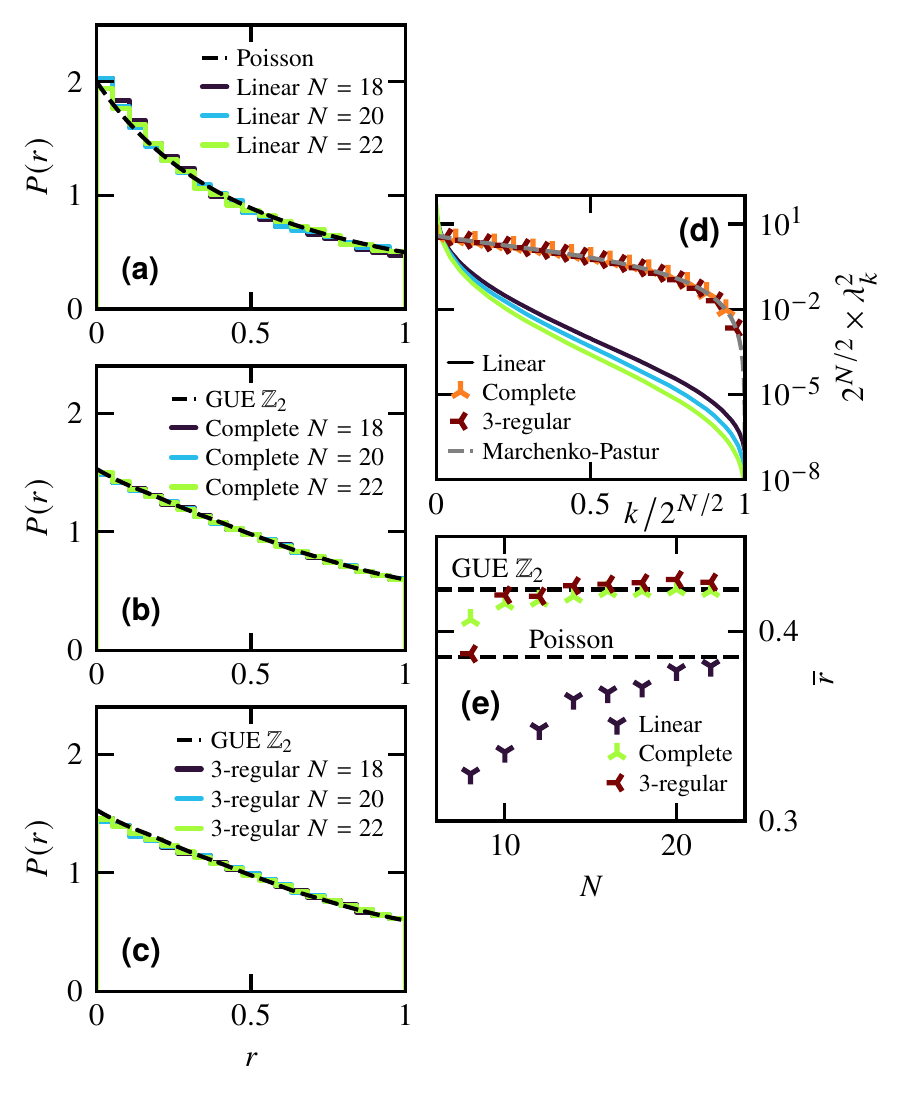} 
    \caption{(a)--(c) Distribution $P(r)$ of the adjacent gap ratio defined in Eq.~\eqref{eq:gapratio} for (a) linear graphs, (b) complete graphs, and (c) $3$-regular graphs of size $N=18$, $20$, and $22$, computed after saturation of the entanglement entropy. (a) The distribution shows a Poisson law, expected for the output of a free-fermionic circuit, see Eq.~\eqref{eq:poisson}. (b), (c) The distribution shows a GUE law with $\mathbb{Z}_2$ symmetry~\cite{PhysRevX.12.011006}. (d) Average entanglement spectrum (note the scaling by $2^{N/2}$ of the $x$ and $y$ axes, corresponding to the number of Schmidt coefficients) for the different graphs considered, computed after saturation of the entanglement entropy. Data averaged over $10^3$ problems and randomized QAOA circuits. The statistical error bars are smaller than the symbols and not displayed. The entanglement spectrum of the complete and $3$-regular graphs are shown for $N=22$ and follow the Marchenko-Pastur of Eq.~\eqref{eq:ent_spec_random_state}. The entanglement spectrum of linear graphs behave differently due to the free-fermionic nature of the QAOA circuit in that case. Finite-size effects are visible (data shown for $N=18$, $20$, and $22$). (e) Average adjacent gap ratio $\overline{r}$ as a function of the system size $N$. The Poisson value $\overline{r}_\mathrm{Poisson}=2\ln 2-1\simeq 0.38629$ and GUE value with $\mathbb{Z}_2$ symmetry $\overline{r}_\mathrm{GUE-\mathbb{Z}_2}=0.422085$~\cite{PhysRevX.12.011006} are also plotted.}
    \label{fig:random_QAOA_ent_spec}
\end{figure}

\subsubsection{Structure of the entanglement spectrum}

The nature of the quantum state following a randomized QAOA circuit can be further characterized through its entanglement spectrum. From a random matrix theory perspective, we expect the entanglement spectrum of a random state to follow the Marchenko-Pastur distribution (up to normalization)~\cite{Marcenko1967,Znidaric2006} once the entanglement entropy reaches saturation. For an equal bipartition in the limit $N\to+\infty$, and rescaling the Schmidt coefficients index $k=0,1,\ldots 2^{N/2}-1$ to the interval $x\in[0,1]$, the Schmidt coefficients become a continuous function of $x$~\cite{Znidaric2006,PhysRevLett.115.267206},
\begin{equation}
    2^{N/2}\lambda^2(x) = 4\cos^2\varphi,~~\mathrm{with}~~\frac{\pi}{2}x=\varphi-\frac{1}{2}\sin\bigl(2\varphi\bigr),
    \label{eq:ent_spec_random_state}
\end{equation}
where the value of $\lambda^2(x)$ can be evaluated numerically.

We analyze in Fig.~\ref{fig:random_QAOA_ent_spec}(d) the entanglement spectrum once entanglement entropy reaches saturation and average the data over $10^3$ problems. $3$-regular and complete graphs follow the Marchenko-Pastur distribution of Eq.~\eqref{eq:ent_spec_random_state} for all sizes $N$ considered (only shown for the largest one, $N=22$). The entanglement spectrum of linear graphs does not follow the Marchenko-Pastur distribution, which is expected as the entanglement entropy does not saturate to the Page value in this case, due to the free-fermionic nature of the circuit. Instead, we see in Fig.~\ref{fig:random_QAOA_ent_spec}(d) that the entanglement spectrum decays much faster for linear graphs, with finite-size effects. To the best of our knowledge, there is no closed-form expression similar to Eq.~\eqref{eq:ent_spec_random_state} in the case of free-fermionic circuits.

\subsubsection{Level statistics of the adjacent gap ratio}

Another tool from random matrix theory to characterize the nature of the entanglement spectrum is the level statistics of the adjacent gap ratio $r$~\cite{PhysRevB.75.155111,PhysRevLett.110.084101},
\begin{equation}
    r_k = \mathrm{max}\bigl(\delta_k, \delta_{k+1}\bigr)\bigr/\mathrm{min}\bigl(\delta_k, \delta_{k+1}\bigr),
    \label{eq:gapratio}
\end{equation}
with $\delta_k=\lambda^2_k-\lambda^2_{k-1}$, where $\{\lambda^2_k\}$ are arranged in ascending order such that $\delta_k\geq 0$ and $r_k\in[0,1]$. The free-fermionic nature of QAOA on linear graphs leads to a Poisson law for the level statistics~\cite{PhysRevB.75.155111,PhysRevLett.110.084101}:
\begin{equation}
    P_\mathrm{Poisson}(r)=2\bigr/\bigl(1+r\bigr)^2.
    \label{eq:poisson}
\end{equation}
The distribution is plotted in Fig.~\ref{fig:random_QAOA_ent_spec}(a), where one observes that $P(r)$ is in perfect agreement with the Poisson distribution of Eq.~\eqref{eq:poisson}. The corresponding average gap ratio $\overline{r}$ is shown in Fig.~\ref{fig:random_QAOA_ent_spec}(d). Despite the slow convergence with system size $N$, the value is compatible with the Poisson value $\overline{r}_\mathrm{Poisson}=2\ln 2-1\simeq 0.38629$ as $N\to+\infty$. 

For $3$-regular and complete graphs, the level statistics of the adjacent gap ratio follows the Gaussian unitary ensemble (GUE). To observe genuine GUE, the $\mathbb{Z}_2$ symmetry of the circuit needs to be considered by performing the Schmidt decomposition and the level statistics analysis on the different blocks independently. For a $\mathbb{Z}_2$ symmetry, the average gap ratio when mixing the two different blocks is $\overline{r}_\mathrm{GUE-\mathbb{Z}_2}=0.422085$, as compared to $\overline{r}_\mathrm{GUE}=0.60266$ for genuine GUE if the blocks were considered separately~\cite{PhysRevX.12.011006}. The distribution is plotted in Figs.~\ref{fig:random_QAOA_ent_spec}(b) and~\ref{fig:random_QAOA_ent_spec}(c), where one observes that $P(r)$ is in perfect agreement with the GUE $\mathbb{Z}_2$ distribution. An analytical expression for $P(r)$ is available in Ref.~\cite{PhysRevX.12.011006}. The average gap ratio as a function of system size $N$ is plotted in Fig.~\ref{fig:random_QAOA_ent_spec}(d), compatible with the expected value. In the case of linear graphs, the mixing of different blocks is not an issue as they both follow a Poisson law, and mixing blocks is known to drive the distribution toward Poisson, independently of their individual nature~\cite{PhysRev.120.1698,PhysRevX.12.011006}.

\section{Entanglement in Optimized QAOA Circuits}
\label{sec:ent_opt}

\begin{figure}[t]
    \includegraphics[width=1\columnwidth]{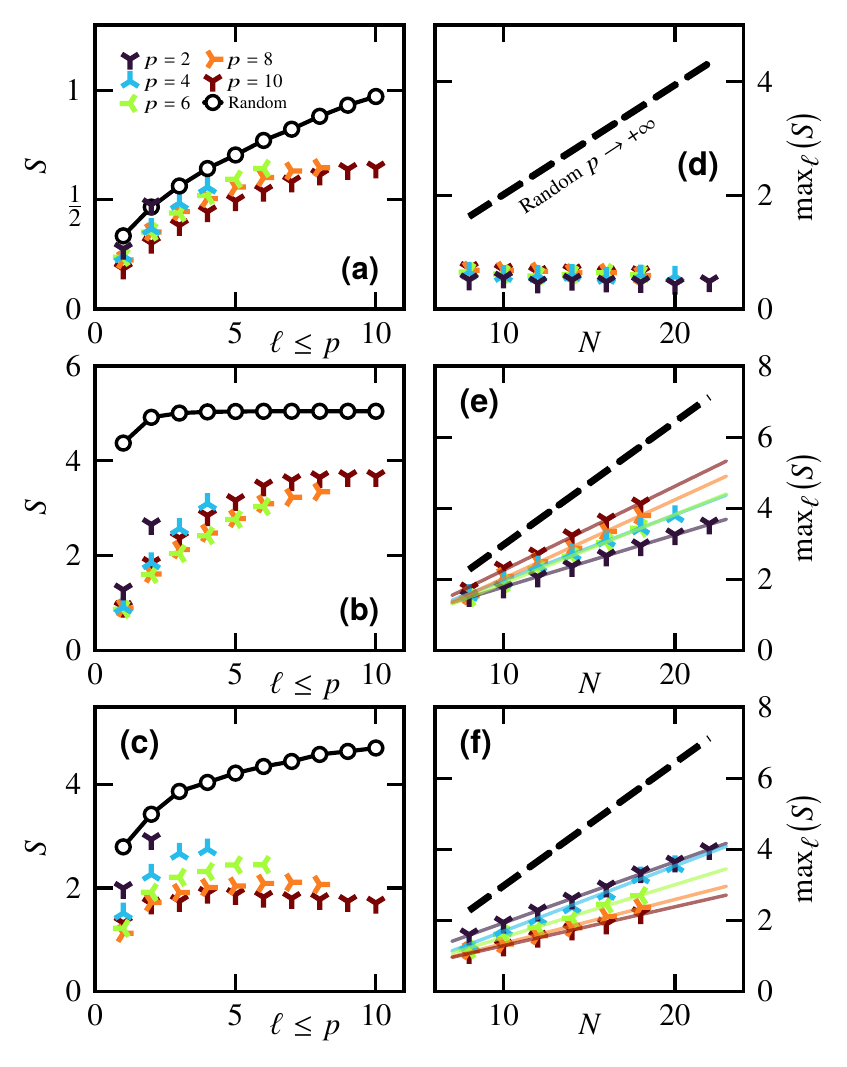} 
    \caption{Data averaged over $10^2$ problems. The statistical error bars are smaller than the symbols and not displayed. (a), (d) Linear graphs. (b), (e) Complete graphs. (c), (f) $3$-regular graphs. Left column: Average bipartite von Neumann entanglement entropy as defined in Eq.~\eqref{eq:entanglement_def} versus the number of optimized QAOA layers $\ell\leq p$ for various QAOA depths $p$. The system size considered is $N=16$. The entanglement in the case of randomized QAOA layer is also displayed. Right column: The data displayed in the left column is generated for various sizes $N$. For each curve $(N,p)$, the maximum value taken by the entropy $S$ as a function of $\ell$ is extracted: It is then plotted on the right column as a function of the system size $N$. The value taken by the entanglement at saturation ($p\to+\infty$) for randomized QAOA circuits, see Eq.~\eqref{eq:entanglement_saturation}, is shown for comparison. Plain lines are linear fits of the form $a(p)+b(p)N$ with $a(p)$ and $b(p)$ fitting parameters.}
    \label{fig:optimized_QAOA_S}
\end{figure}

We now investigate the entanglement generated by optimized QAOA circuits. For a given problem, an optimized QAOA circuit is obtained by running $10^3$ simulations with random initial parameters $\boldsymbol{\theta}$ and only keeping the instance resulting in the lowest cost.

In a QAOA circuit, see Eq.~\eqref{eq:qaoa_circuit}, the initial state $H^{\otimes N}\vert 0\rangle^{\otimes N}$ before applying the parametric gates leads to a product state with therefore no entanglement: $S\bigl(\ell=0\bigr)=0$. Assuming the solution to a given MaxCut problem is nondegenerate, an optimized QAOA circuit in the limit of large depth will output a cat state of the form $\bigl(\vert{\boldsymbol{s}_\mathrm{exact}}\rangle+\mathrm{e}^{i\phi}\vert{-\boldsymbol{s}_\mathrm{exact}}\rangle\bigr)/\sqrt{2}$, with $\phi$ an angle (up to an irrelevant global phase). This state has an entanglement $S=\ln 2$, independently of the number of qubits involved. The two extremum states have little or no entanglement, as compared to the randomized case of the previous section where $S\sim N$. Both states can be exactly represented with MPSs with a bond dimension $\chi=2$. Therefore, it is legitimate to investigate what happens in between the initial and final layers by considering optimized QAOA circuits.

\subsection{Entanglement Entropy}

\subsubsection{Entanglement growth}

We first set the system size to $N=16$ and see how the bipartite von Neumann entanglement grows after each optimized QAOA layer $\ell\leq p$ for various QAOA depths $p\leq 10$. The results are plotted in Figs.~\ref{fig:optimized_QAOA_S}(a)--(c) for linear, complete, and $3$-regular graphs, respectively.

A first observation is that, in all cases, the entanglement growth in optimized QAOA circuits is slower than in randomized circuits. A second observation is that on the linear and $3$-regular graphs, increasing the QAOA depth $p$ reduces the amount of entanglement produced by each layer $\ell\leq p$. This observation does not hold true for the complete graphs, and we attribute this difference to the extensive connectivity of each vertex $\sim N$, which is constant for linear (two) and $3$-regular graphs (three). Note that for $p=10$ in the $3$-regular graph case, we see that the maximum of entanglement happens after an intermediate layer $\ell=4$, before decreasing, in line with the sketch of Fig.~\ref{fig:introduction}(a).

\subsubsection{Maximum of entanglement}

We repeat the previous procedure for various system sizes $N\leq 22$. For each curve $(N,p)$, the maximum value taken by the entropy $S$ as a function of $\ell$ is extracted. The value $\mathrm{max}_\ell(S)$ is plotted versus $N$ for various QAOA depths $p$ in Figs.~\ref{fig:optimized_QAOA_S}(d)--(f) for linear, complete, and $3$-regular graphs, respectively.

\paragraph{$3$-regular and complete graphs.} The maximum of entanglement generated by an optimized QAOA circuits follows a volume law with $\mathrm{max}_\ell(S)=a(p)+b(p)N$, with $a(p)$ and $b(p)$ parameters. For the cases $p\leq 10$ considered, the slope is smaller than for randomized QAOA circuits for which $b(p\to+\infty)=\ln 2/2$, see Eq.~\eqref{eq:entanglement_saturation}. For the $3$-regular graphs, the observation that increasing the QAOA depth $p$ reduces the amount of entanglement produced by each layer $\ell\leq p$ is also observed in Fig.~\ref{fig:optimized_QAOA_S}(f), with the slope $b(p)$ decreasing with increasing $p$.

\paragraph{Linear graphs.} The behavior is totally different, see Fig.~\ref{fig:optimized_QAOA_S}(d): At fixed QAOA depth $p$, the maximum of entanglement is roughly constant as a function of the system size. This was also the case in randomized QAOA circuits of Fig.~\ref{fig:random_QAOA_S}(a), where the entanglement growth was diffusive $\propto\sqrt{p}$ with a size-independent prefactor. There, as long as $p\lesssim N^2$---which is verified even for small system sizes since practical values of $p$ are unlikely to scale as the system size square---the entanglement shows no size dependence.

\subsection{Entanglement Spectrum}

\begin{figure}[t]
    \includegraphics[width=1\columnwidth]{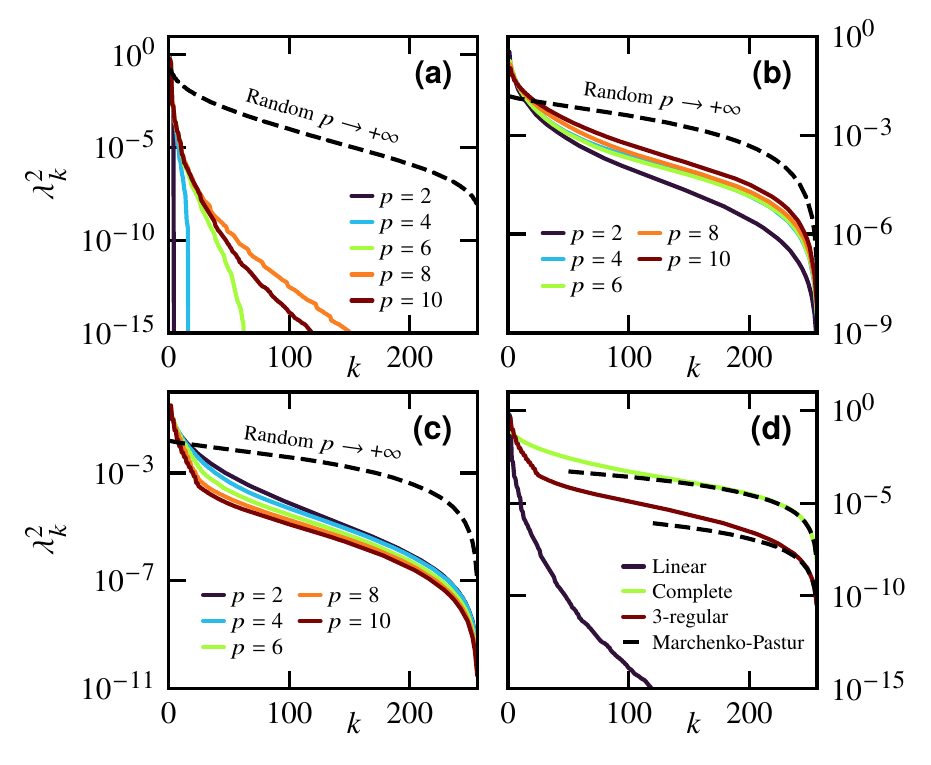} 
    \caption{Data averaged over $10^2$ problems. The statistical error bars are smaller than the symbols and not displayed. (a) (b) (c) Average entanglement spectrum for $N=16$ computed at the end of optimized QAOA circuits ($p=\ell$) for various QAOA depths $p$. The average entanglement spectrum in the randomized QAOA case for $N=16$ computed once the entanglement entropy has reached its saturation value is also displayed. (a) Linear graphs. (b) Complete graphs. (c) $3$-regular graphs. (d) Average entanglement spectrum for $N=16$ and $p=\ell=10$ for the three different graphs considered in this paper. The dashed line corresponds to the Marchenko-Pastur distribution of Eq.~\eqref{eq:ent_spec_random_state}, where the curve has been vertically adjusted to fall over the QAOA data.}
    \label{fig:optimized_QAOA_ent_spec}
\end{figure}

\subsubsection{Decay of the entanglement spectrum}

We now turn our attention to the entanglement spectrum generated by optimized QAOA circuits. We compute the entanglement spectrum at the end of the circuit ($p=\ell$) for $N=16$ and various values of QAOA depths $p$, see Figs.~\ref{fig:optimized_QAOA_ent_spec}(a)--(c) for linear, complete, and $3$-regular graphs, respectively.

The most noticeable difference between the three cases concerns linear graphs. Here, the entanglement spectrum decays extremely fast, with only a finite fraction of them entering the computation of the bipartite von Neumann entanglement entropy via Eq.~\eqref{eq:entanglement_def}. Because of the topology of linear graphs, where the entanglement grows at most linearly with the circuit depth $p$ in a light cone fashion, we do not expect the results to change substantially for larger system sizes: Qubits outside of a region of length $p$ cannot be entangled, independently of the system size $N$.

In the case of complete and $3$-regular graphs, there is no such strong decay for $\lambda^2_k$ as a function of $k$. Nevertheless, when compared to the random case, the first few $\lambda^2_k$ decay much faster, over several order of magnitudes. This is especially visible for $3$-regular graphs in Fig.~\ref{fig:optimized_QAOA_ent_spec}(c) where the decay for small $k$ gets sharper as $p$ increases. The opposite behavior is observed for complete graphs: The decay rate gets smaller as $p$ increases. However, this can only be a transient effect since the $p\to+\infty$ limit should give the unique solution to the MaxCut problem, with only two nonzero values $\lambda_0^2=\lambda_1^2=0.5$, giving the bipartite von Neumann entropy $S=\ln 2$.

\subsubsection{Two-component structure}

We compare in Fig.~\ref{fig:optimized_QAOA_ent_spec}(d) the average entanglement spectrum for $N=16$ and $p=\ell=10$ for the three graphs. The plot emphasizes the fast decay of linear graphs. For the complete and $3$-regular graphs, we find that the tail of the entanglement spectrum follows the Marchenko-Pastur distribution of Eq.~\eqref{eq:ent_spec_random_state}, highlighting two different components in the spectrum at small and large $k$. As the Marchenko-Pastur distribution arises for random states, it reveals that the state resulting from an optimized QAOA circuit at finite $p$ still carries random features. This is especially true for complete graphs, where the overlap between the data and Eq.~\eqref{eq:ent_spec_random_state} holds for $k/2^{N/2}\gtrsim 0.5$ while it only holds for $k/2^{N/2}\gtrsim 0.9$ in the case of $3$-regular graphs. A similar two-component behavior was reported~\cite{PhysRevLett.115.267206} in highly excited eigenstates of disordered one-dimensional Hamiltonians satisfying the eigenstate thermalization hypothesis~\cite{PhysRevA.43.2046,PhysRevE.50.888,Rigol2008} (in the context of the many-body localization transition). Although this problem and QAOA are unrelated, a parallel can be drawn: There needs to be enough randomness to produce a volume law scaling of the entanglement---albeit with a rate lower than the Page value of Eq.~\eqref{eq:entanglement_saturation}---, and yet both problems have built-in processes that favor certain eigenvalues $\lambda_k^2$ of the reduced density matrix, i.e., the solution to the MaxCut problem versus the disorder configuration. The entanglement scaling in the context of optimized QAOA circuits is reported in Figs.~\ref{fig:optimized_QAOA_S}(e) and~\ref{fig:optimized_QAOA_S}(f).

\subsection{The Special Case of Linear Graphs}

We emphasized the unique nature of QAOA circuits for linear graphs which map to free fermionic circuits. It has been shown recently that arbitrary free fermionic circuits can be compressed without loss in polynomial time (with respect to their depth and number of qubits) such that their final depth is at most linear in the number of qubits~\cite{Kokcu2021a,Kokcu2021b,Camps2021}. In particular, QAOA circuits for $N$ qubits defined in Eq.~\eqref{eq:qaoa_circuit} can be compressed such that at most $p=N$. This means that for linear graphs, there exists a QAOA circuit of depth at least $p=N$ solving the MaxCut problem exactly.

\section{Entanglement in a Quantum Annealing Protocol}
\label{sec:ent_annealing}

QAOA and quantum annealing seek to solve the same class of optimization problems from a different perspective. Quantum annealing relies on quantum adiabaticity. One first prepares a system in the ground state of $-\sum_{n=1}^NX_n$, and then interpolates over a period of time $T$ between this initial Hamiltonian and the final one, i.e., the cost function $C$ of which we want to find the ground state. Here, $C$ is defined in Eq.~\eqref{eq:qaoa_cost}. Formally the interpolation Hamiltonian can be written as,
\begin{equation}
    \mathcal{H}\bigl(T,t\bigr) = -\Bigl(1-\frac{t}{T}\Bigr)\sum_{n=1}^NX_n + \frac{t}{T}C,~\mathrm{with}~t\in[0,T],
    \label{eq:ham_interpolate}
\end{equation}
with $t$ the time. In the adiabatic limit, i.e., $T\to+\infty$, the time evolution,
\begin{equation}
    \vert t\rangle=\mathcal{T}\mathrm{exp}\left[-i\int^{t}_0dt'\,\mathcal{H}\bigl(T,t'\bigr)\right]H^{\otimes N}\vert 0\rangle^{\otimes N},
    \label{eq:time_evolution}
\end{equation}
will give a state $\vert t=T\rangle$ corresponding to the ground state of $C$ defined in Eq.~\eqref{eq:qaoa_cost}. $\mathcal{T}$ indicates a time-ordered exponential. Eq.~\eqref{eq:time_evolution} can be discretized by introducing a finite time-step $\delta t$ along with a Trotterization---which in the limit $\delta t\to 0$ gives back the continuous form of Eq.~\eqref{eq:time_evolution}:
\begin{equation}
    \vert t\rangle=\Biggr(\prod_{\ell=1}^{t/\delta t}U_{\beta_\ell}U_{\gamma_\ell}\Biggl)H^{\otimes N}\vert 0\rangle^{\otimes N}.
    \label{eq:time_evolution_discretized}
\end{equation}
The resulting quantum circuit is analogous to QAOA of Eq.~\eqref{eq:qaoa_circuit} with the unitaries defined in Eq.~\eqref{eq:qaoa_unitaries}. The equivalent of QAOA depth is $T\delta t^{-1}$ ($\equiv p$) and the parameters in each layer $\ell$, evolving the state by a time step $\delta t$, are now simply given by the interpolation of Eq.~\eqref{eq:ham_interpolate}, i.e., $\gamma_\ell=2t\delta t/T$ and $\beta_\ell=-2\delta t(1-t/T)$. There is no hybrid quantum-classical optimization involved. Adiabatic computation requires that the total time evolution scales as $T\sim\Delta^{-2}$ with $\Delta$ the minimum spectral gap~\cite{RevModPhys.90.015002}. $\Delta$ is a property of the problem of interest, with no easy way to determine it a priori.

We investigate the entanglement properties of the quantum annealing circuit of the discretized version of Eq.~\eqref{eq:time_evolution}, using a time step $\delta t=0.1$.

\begin{figure}[t]
    \includegraphics[width=1\columnwidth]{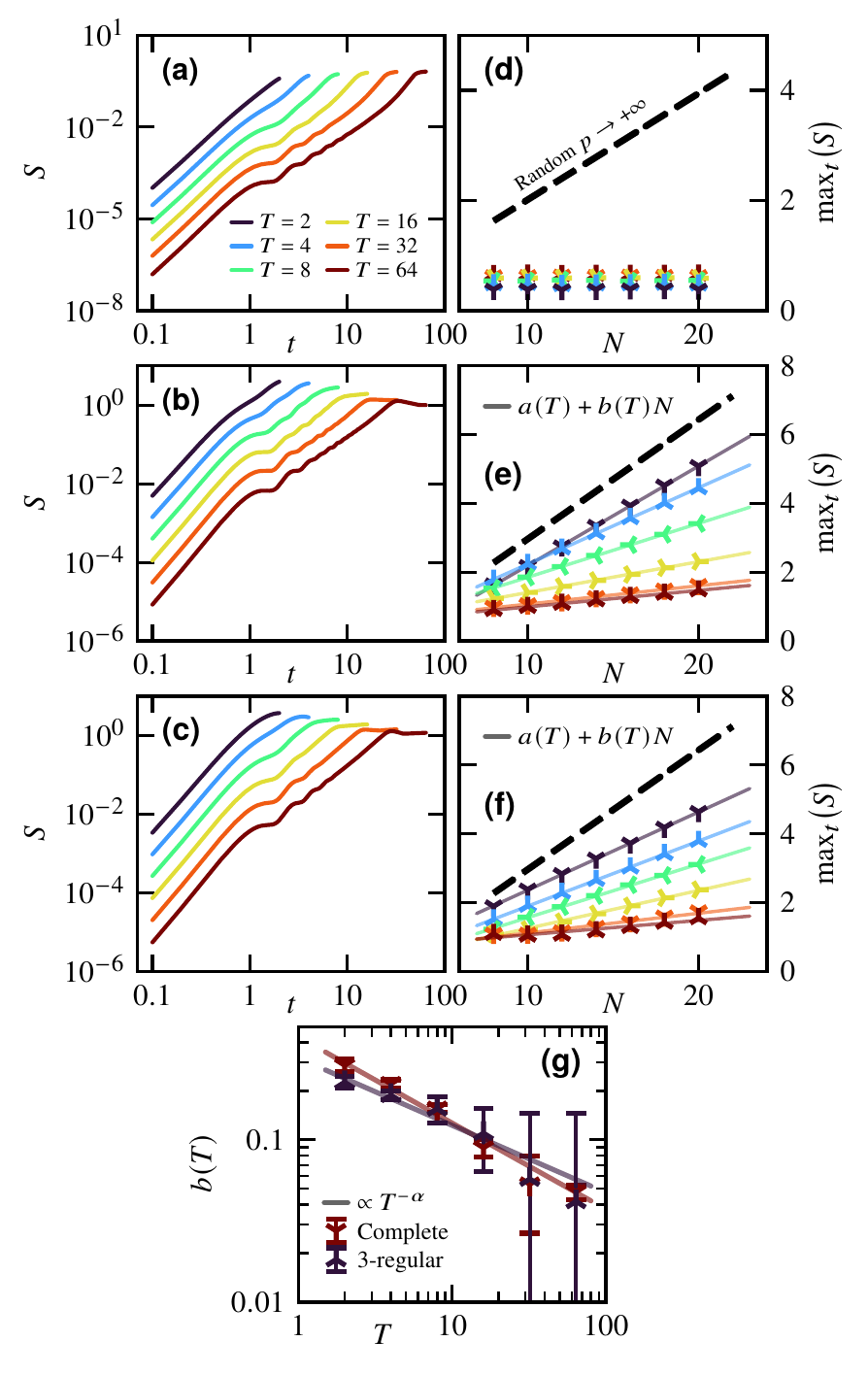} 
    \caption{Data averaged over $10^3$ problems. The statistical error bars are smaller than the symbols and not displayed. (a), (d) Linear graphs. (b), (e) Complete graphs. (c), (f) $3$-regular graphs. Left column: Average bipartite von Neumann entanglement entropy for $N=16$ computed as a function of the time $t$ for various total evolution times $T$ following Eqs.~\eqref{eq:ham_interpolate} and~\eqref{eq:time_evolution_discretized}. Right column: The data displayed in the left column is generated for various sizes $N$. For each curve $(N,T)$, the maximum value taken by the entropy $S$ as a function of $t$ is extracted: It is then plotted on the right column as a function of the system size $N$. The value taken by the entanglement at saturation for randomized QAOA circuits ($p\to+\infty$), see Eq.~\eqref{eq:entanglement_saturation}, is shown for comparison. Plain lines are linear fits of the form $a(T)+b(T)N$ with $a(T)$ and $b(T)$ fitting parameters. (g) Fitting parameter $b(T)$ for $3$-regular and complete graphs showing an algebraic decay $\propto T^{-\alpha}$ with $\alpha\approx 0.5$.}
    \label{fig:adiabatic_S}
\end{figure}

\subsection{Entanglement growth}

First, at fixed system size $N=16$, we compute the average bipartite von Neumann entanglement entropy as a function of the time $t$ for various values of total evolution times $T$ following Eqs.~\eqref{eq:ham_interpolate} and~\eqref{eq:time_evolution_discretized}. The data is shown in Figs.~\ref{fig:adiabatic_S}(a)--(c) for linear, complete, and $3$-regular graphs, respectively. At $t=0$, the state is a product state, thus unentangled with $S=0$. At finite but small time $t$, the entanglement grows algebraically as $\propto t^\kappa$ with $\kappa\approx 2.7$ in the three cases, independently of the total evolution time $T$. The entanglement then takes a maximum value which, for regular random and complete graphs, decrease as $T$ increases. This may suggest that in the adiabatic limit with $T\to+\infty$, the entanglement generated by the time evolution will not exceed the entanglement of the exact solution, i.e., $S=\ln 2$ for MaxCut problems with a unique solution. For linear graphs, this observation seems to hold independently of $T$.

\subsection{Maximum of entanglement}

In Figs.~\ref{fig:adiabatic_S}(d)--(f), we repeat the previous analysis for different system sizes $N$ and consider for each curve $(N,T)$ the maximum of entanglement reached during the time evolution $t\in[0,T]$. The results are plotted versus $N$ for various total evolution times $T$. For linear graphs, we find that the maximum of entanglement is independent of $N$. On the other hand, for $3$-regular and complete graphs, the maximum of entanglement shows a volume law with a linear scaling of the form $S=a(T)+b(T)N$ with $a(T)$ and $b(T)$ fitting parameters. The fitting parameter $b(T)$ is plotted in Fig.~\ref{fig:adiabatic_S}(g). Its behavior is compatible with an algebraic decay $\propto T^{-\alpha}$ with $\alpha\approx 0.5$.

\section{Summary and implications}
\label{sec:conclusion}

In this paper, we investigated the entanglement growth and spread generated by randomized and optimized QAOA circuits for solving MaxCut problems on different types of graphs. We also considered the entanglement spectrum in connection with random matrix theory. In addition, we studied entanglement production in a quantum annealing protocol aiming to solve the same optimization problems.

In the latter case, we found that for unit-weight $3$-regular and random-weight complete graphs, the maximum of entanglement grows as $S\sim NT^{-\alpha}$ with $T$ the total evolution time and $\alpha\approx 0.5$. For a given problem, a polynomial scaling of the minimum spectral gap with the system size $\Delta\sim N^{-\beta}$ with $\beta>0$ would imply that $T$ should also scale polynomially with $N$ for the time evolution to be adiabatic, i.e., $T\sim N^{2\beta}$ versus $T\sim\exp(2\beta N)$ for an exponential scaling. The entanglement scaling with $N$ is favorable in the exponential case, but it requires exponentially more time---layers of gates---for the evolution to remain adiabatic. In the polynomial case, the entanglement scaling with $N$ is favorable if $2\beta\alpha\geq 1$, meaning that one can simulate the circuit with MPS using a finite bond dimension $\chi$ and a polynomial number of layers. The scaling of the minimum spectral gap with the system size is problem dependent. See, e.g., Ref.~\cite{BAPST2013127} and references therein for a review on quantum annealing applied to combinatorial optimization problems.

QAOA attempts to circumvent the extensive circuit depth required by an adiabatic time evolution by optimizing the circuit parameters instead. We found that both randomized and optimized QAOA circuits generate volume law entanglement $S\sim N$. Although the prefactor is smaller in the optimized case, suggesting that a smaller bond dimension $\chi$ would be required to simulate optimized QAOA circuits versus random ones, it does not help much as one still has to find the best QAOA angles. Indeed, a typical starting point is random angles, and if one were able to do better than that initially, optimization algorithms need to explore the cost landscape to find its minimum, meaning that the quantum state may need to accommodate larger entanglement during intermediate steps.

We noted that linear graphs are special due to the mapping of QAOA circuits to free fermionic circuits which can be simulated in polynomial time. Moreover, such circuits can be compressed such that their final depth is at most linear in the number of qubits~\cite{Kokcu2021a,Kokcu2021b,Camps2021}.

In summary, we found that there is an entanglement barrier $S\sim N$ to cross for large-depth QAOA circuits to go from an initial product state $S=0$ to a final state with low-entanglement solving the optimization problem of interest. This barrier makes it difficult for entanglement-based simulation methods like MPSs to execute accurately QAOA circuits with a finite bond dimension $\chi$~\cite{Dupont2022}.

\textit{Note added:} Recently, we became aware of Ref.~\cite{Chen2022,Sreedhar2022} also investigating entanglement in QAOA circuits.

\begin{acknowledgments}
    We acknowledge discussions with S. Cohen, J-S. Kim, and S. Stanwyck at Nvidia during the early stages of this work. This work was supported by the U.S. Department of Energy (DOE), Office of Science, National Quantum Information Science Research Centers (NQISRC), Superconducting Quantum Materials and Systems Center (SQMS) under contract No. DE-AC02-07CH11359, and by the Quantum Science Center (QSC), a National Quantum Information Science Research Center of the DOE. J.E.M. was also supported by a Simons Investigatorship. This research used the Lawrencium computational cluster resource provided by the IT Division at the Lawrence Berkeley National Laboratory (supported by the Director, Office of Science, Office of Basic Energy Sciences, of the U.S. Department of Energy under Award No. DE-AC02-05CH11231). This research also used resources of the National Energy Research Scientific Computing Center, a DOE Office of Science User Facility supported by the Office of Science of the U.S. Department of Energy under Contract No. DE-AC02-05CH11231 using NERSC Award No. DDR-ERCAP0022242.
\end{acknowledgments}

\bibliography{references}

\end{document}